\input{amssymb.sty}
\documentstyle[aps,prl,floats,eqsecnum,epsf,epsfig]{revtex}

\input{psfig.sty}
\begin{document}
\draft

\twocolumn[\hsize\textwidth\columnwidth\hsize\csname@twocolumnfalse\endcsname 

\title{Crossover critical behavior in the three-dimensional Ising model}
\author{Young C. Kim$^{1}$, Mikhail A. Anisimov$^{1,2}$, Jan V. Sengers$^{1,2}$, and Erik Luijten$^{3}$}
\address{$^{1}$Institute for Physical Science and Technology, University of Maryland,\\ College Park, MD 20742\\
$^{2}$Department of Chemical Engineering, University of Maryland,\\ College Park, MD 20742\\
$^{3}$Department of Materials Science and Engineering, University of Illinois, \\ Urbana, IL 61801}

\date{\today}
\maketitle
\begin{abstract}
The character of critical behavior in physical systems depends on the range of interactions. In the limit of infinite range of the interactions, systems will exhibit mean-field critical behavior, i.e., critical behavior not affected by fluctuations of the order parameter. If the interaction range is finite, the critical behavior asymptotically close to the critical point is determined by fluctuations and the actual critical behavior depends on the particular universality class. A variety of systems, including fluids and anisotropic ferromagnets, belongs to the three-dimensional Ising universality class. Recent numerical studies of Ising models with different interaction ranges have revealed a spectacular crossover between the asymptotic fluctuation-induced critical behavior and mean-field-type critical behavior. In this work, we compare these numerical results with a crossover Landau model based on renormalization-group matching. For this purpose we consider an application of the crossover Landau model to the three-dimensional Ising model without fitting to any adjustable parameters. The crossover behavior of the critical susceptibility and of the order parameter is analyzed over a broad range (ten orders) of the scaled distance to the critical temperature. The dependence of the coupling constant on the interaction range, governing the crossover critical behavior, is discussed
\end{abstract}
\vspace{.3in}
]

\vspace{.2in}
\section{Introduction}
In systems with a critical-point phase transition the asymptotic critical behavior of thermodynamic properties can be characterized by scaling laws with universal critical exponents and universal scaling functions \cite{ref1}. In this issue of the Journal of Statistical Physics, dedicated to M. E. Fisher, it is interesting to recall some of the discussions at a conference on critical phenomena held in Washington, DC in 1965 \cite{ref2}. At that conference, Fisher reviewed the definitions of the proposed asymptotic critical power laws, both for the thermodynamic properties \cite{ref3} and for the correlation function of the order parameter \cite{ref4}. Commenting on the presentations of Fisher, Debye \cite{ref5} made the following remark: ``I would like that the theoretical people tell me when I am so and so far away from the critical point, then my curve should look so and so,'' to which Fisher responded \cite{ref6}: ``What Dr. Debye said is correct, namely that when one is talking about some deviations one must attach to it a magnitude and a range of temperatures where it is to be observed, which can be different for one phenomenon compared to another. When these ranges differ from one system to another for the same phenomenon, we have a real challenge.'' It is the purpose of the present paper to further elucidate the issue of nonasymptotic critical behavior.

The asymptotic fluctuation-induced critical scaling laws are valid sufficiently close to the critical point, where the correlation length $\xi$ of the long-range critical fluctuations will be much larger than the range of the intermolecular interaction. Away from the critical point, the correlation length $\xi$ decreases and sufficiently far away from the critical point $\xi$ will become of the order of the intermolecular interaction range and the effects of any critical fluctuations become negligible.

The actual critical behavior will depend strongly on the range of the microscopic interactions. As the interaction range becomes larger and larger, the critical fluctuations become more and more suppressed and mean-field theory will become applicable up to temperatures closer and closer to the critical temperature. In the limit of infinite interaction one recovers classical, i.e., mean-field asymptotic critical behavior. Hence, as the interaction range varies, the system will exhibit a crossover from fluctuation-induced critical behavior to mean-field critical behavior.

Recently, Luijten and coworkers have reported numerical studies of the susceptibility and the order parameter of both two-dimensional \cite{ref7,ref8,ref9} and three-dimensional \cite{ref10,ref11,ref12} Ising models with different interaction ranges, showing crossover from fluctuation-induced to mean-field critical behavior. On the other hand, during the past two decades theoretical equations for dealing with crossover from fluctuation-induced to mean-field critical behavior have been developed by several groups of investigators \cite{ref13,ref14,ref15,ref16,ref17,ref18,ref19,ref20,ref21,ref22,ref23,ref24,ref25,ref26,ref27,ref28,ref29}. A comparison between some of these theoretical approaches \cite{ref16,ref17,ref18,ref19,ref20,ref21,ref24} has been presented by Anisimov {\em et al.} \cite{ref22}. Here we consider a crossover Landau model that has been used to represent experimental thermodynamic-property data in the critical region of fluids, which are expected to belong to the three-dimensional Ising universality class \cite{ref20,ref21,ref30,ref31}. The crossover Landau model contains some system-dependent parameters related to the coefficients in a Landau-Ginzburg Hamiltonian \cite{ref22}. Since these system-dependent coefficients are not known a priori for fluids, they are used as adjustable parameters in fits to experimental data. However, in the case of the Ising models studied numerically by Luijten and Binder \cite{ref10,ref11} we have a priori information concerning the system-dependent constants. Hence, it has become possible to evaluate the crossover Landau model for Ising models without using adjustable parameters and to make a comparison with the numerical data obtained by Luijten and Binder for three-dimensional Ising models.

\section{Crossover Landau Model}
A procedure for constructing an expression for the Helmholtz-energy density for fluids in the critical region, that incorporates crossover from Ising-like critical behavior to classical behavior was originally proposed by Chen and coworkers \cite{ref19,ref20,ref21}. The procedure, based on a theoretical analysis of Nicoll and coworkers \cite{ref13,ref14,ref15}, represents an approximate solution of the nonlinear renormalization-group equations and involves a transformation to a classical (Landau) expansion. Two crossover Landau models, a two-term and a six-term crossover Landau model have been considered depending on the number of terms retained in the classical Landau expansion as further reviewed by Anisimov {\em et al.} \cite{ref22}. Explicit expressions for the amplitudes of the critical power laws incorporated in the crossover Landau models have been evaluated by Tang {\em et al.} \cite{ref32} and, more recently, by Agayan {\em et al.} \cite{ref33}. Here we consider the two-term crossover model which deals with the crossover from asymptotic Ising-like critical behavior to asymptotic mean-field critical behavior \cite{ref22}.

In the crossover Landau model one starts with an asymptotic Landau expansion of the critical part $\Delta A$ of the classical local Helmholtz free-energy density $A$ \cite{ref22}:
 \begin{equation}
  \Delta A = \mbox{$\frac{1}{2}$}a_{0} \tau \phi^{2} + \mbox{$\frac{1}{4!}$}u_{0}\phi^{4} + \mbox{$\frac{1}{2}$}c_{0}(\nabla\phi)^{2},  \label{class}
 \end{equation}
where $\tau = (T-T_{\mbox{\scriptsize c}})/T$ is the reduced difference between the actual temperature $T$ and the critical temperature $T_{\mbox{\scriptsize c}}$, $\phi = \phi({\bf r})$ is a spatially fluctuating order parameter, while $a_{0}$, $u_{0}$ and $c_{0}$ are system-dependent coefficients. For the Ising model the order parameter $\phi$ corresponds to the magnetization density. It is convenient to transform the variables and coefficients, such that \cite{ref33}
 \begin{eqnarray}
  M & = & c_{\rho}\phi, \hspace{.2in} t = c_{t}\tau, \hspace{.2in} \tilde{\nabla} = (v_{0}^{1/3}/\pi) \nabla, \\
  a_{0} & = & c_{\rho}^{2}c_{t}, \hspace{.2in} u_{0} = u^{\ast}\bar{u}\Lambda c_{\rho}^{4}, \hspace{.2in} c_{0} = c_{\rho}^{2}v_{0}^{2/3}/\pi^{2},  \label{coeff}
 \end{eqnarray}
where $v_{0}$ is the volume of an elementary lattice cell, $u^{\ast} \simeq 0.472$ is the universal coupling constant at the Ising fixed point \cite{ref32,ref34}, $\bar{u}$ is the scaled $\varphi^{4}$ coupling constant, $\Lambda$ is a dimensionless cut-off wave number, while $c_{t}$ and $c_{\rho}$ are amplitudes of the rescaled temperature scaling field and of the rescaled magnetization density, respectively. Expression (2.1) for $\Delta A$ can then be written as
 \begin{equation}
   \Delta A = \mbox{$\frac{1}{2}$} t M^{2} + \mbox{$\frac{1}{4!}$}u^{\ast}\bar{u}\Lambda M^{4} + \mbox{$\frac{1}{2}$}(\tilde{\nabla} M)^{2}.
 \end{equation}

Upon applying a transformation derived from renorma-\\lization-group matching \cite{ref20,ref21} one obtains a renormalized critical part of the Helmholtz-energy density:
 \begin{equation}
  \Delta A_{\mbox{\scriptsize s}} = \mbox{$\frac{1}{2}$}tM^{2}{\cal T}{\cal D} + \frac{\bar{u}u^{\ast}\Lambda}{4!}M^{4}{\cal D}^{2}{\cal U} -\mbox{$\frac{1}{2}$}t^{2}{\cal K},
 \end{equation}
where, in practice, the following approximate expressions for the rescaling functions ${\cal T}$, ${\cal D}$, ${\cal U}$, and ${\cal K}$ are used:
 \begin{eqnarray}
  {\cal T} & = & Y^{(2\nu -1)/\Delta_{\mbox{\tiny s}}},  \\
  {\cal D} & = & Y^{-\eta\nu/\Delta_{\mbox{\tiny s}}},  \\
  {\cal U} & = & Y^{\nu/\Delta_{\mbox{\tiny s}}}, 
 \end{eqnarray}
and
 \begin{equation}
  {\cal K}  =  \frac{\nu}{\alpha\bar{u}\Lambda}(Y^{-\alpha/\Delta_{\mbox{\tiny s}}} - 1).
 \end{equation}
Here $\nu \simeq 0.630$, $\eta \simeq 0.033$, and $\alpha \simeq 0.110$ are the critical exponents associated with the correlation length, the spatial dependence of the order-parameter at the critical point and the heat capacity, such that $d\nu = 2-\alpha$, where $d=3$ is the dimensionality \cite{ref35,ref36,ref37}, while $\Delta_{\mbox{\scriptsize s}} \simeq 0.51$ is a correction-to-scaling exponent \cite{ref35,ref38}. In the asymptotic crossover theory, the crossover function $Y$ is to be evaluated from the equation \cite{ref22}
 \begin{equation}
  1 - (1-\bar{u})Y = \left(\frac{\bar{u}\Lambda}{\kappa}\right)Y^{\nu/\Delta_{\mbox{\tiny s}}},
 \end{equation}
where the parameter $\kappa$ plays the role of the distance from the critical point and it is given by
 \begin{equation}
  \kappa^{2} = t{\cal T} + \mbox{$\frac{1}{2}$}\bar{u}u^{\ast}\Lambda M^{2}{\cal D}{\cal U}.
 \end{equation}

In zero ordering field, $M=0$, the isothermal susceptibility $\chi \left[ \chi^{-1} = (\partial^{2} A/ \partial \phi^{2})_{\tau} = \right. $ $ \left.(\partial^{2} A/ \partial M^{2})_{t} c_{\rho}^{-2} \right]$ above $T_{\mbox{\scriptsize c}}$ has an expansion of the form
 \begin{equation}
  \chi = \Gamma_{0}^{+}t^{-\gamma}(1+\Gamma_{1}^{+}t^{\Delta_{\mbox{\tiny s}}} + \cdots), \label{sus1}
 \end{equation}
and along the line of spontaneous magnetization (coexistence curve) $M=M_{\mbox{\scriptsize cxc}}$, below $T_{\mbox{\scriptsize c}} (\tau < 0)$:
 \begin{equation}
  \chi = \Gamma_{0}^{-}|t|^{-\gamma}(1+\Gamma_{1}^{-}|t|^{\Delta_{\mbox{\tiny s}}} + \cdots).
 \end{equation}
The order parameter $\phi_{\mbox{\scriptsize cxc}}$ along the ``coexistence curve'' below the critical temperature $(\tau < 0)$ has the expansion
 \begin{equation}
  \phi_{\mbox{\scriptsize cxc}} = c_{\rho}^{-1}M_{\mbox{\scriptsize cxc}} = \pm B_{0}|t|^{\beta}(1 + B_{1}|t|^{\Delta_{\mbox{\tiny s}}} + \cdots ).
 \end{equation}
In the above equations, $\gamma = (2-\eta)\nu \simeq 1.239$ and $\beta = (d\nu - \gamma)/2 \simeq 0.326$ are again universal critical exponents \cite{ref35,ref36,ref37}, while $\Gamma_{0}^{\pm}$, $\Gamma_{1}^{\pm}$, $B_{0}$ and $B_{1}$ are system-dependent amplitudes. These system-dependent amplitudes are related to $\bar{u}\Lambda$ and the coefficients in the transformations (2.2) and (2.3) and they are given by the following equations \cite{ref32,ref33,ref39}. For the leading amplitudes one has
 \begin{eqnarray}
  \Gamma_{0}^{+} & = & g^{+}\left(\frac{\sqrt{c_{t}}}{\bar{u}\Lambda}\right)^{2(1-\gamma)}a_{0}^{-1}, \\
  \Gamma_{0}^{-} & = & g^{-}\left(\frac{\sqrt{c_{t}}}{\bar{u}\Lambda}\right)^{2(1-\gamma)}a_{0}^{-1}, \\
  B_{0} & = & b \left(\frac{\sqrt{c_{t}}}{\bar{u}\Lambda}\right)^{2(\beta-1/2)}(a_{0}/u_{0})^{1/2},
 \end{eqnarray}
and for the correction-to-scaling amplitudes
 \begin{eqnarray}
  \Gamma_{1}^{+} & = & g_{1}^{+} \left(\frac{\sqrt{c_{t}}}{\bar{u}\Lambda}\right)^{2\Delta_{\mbox{\tiny s}}}(1-\bar{u}),  \label{sus2} \\
  B_{1} & = & b_{1}\left(\frac{\sqrt{c_{t}}}{\bar{u}\Lambda}\right)^{2\Delta_{\mbox{\tiny s}}}(1-\bar{u}).
 \end{eqnarray}
The coefficients $g^{\pm}$, $b$, $g_{1}^{+}$, and $b_{1}$ in the above equation have been evaluated by Tang {\em et al.} \cite{ref32}.\footnote{In Ref. \cite{ref32} Tang {\em et al.} actually considered a variety of approximants for the rescaling functions ${\cal T}$, ${\cal D}$, ${\cal U}$, and ${\cal K}$. The crossover Landau model considered here corresponds to the one designated by Tang {\em et al.} as crossover model II.} They depend on the critical exponents and the coupling constant $u^{\ast}$ and are therefore again universal constants with the values \cite{ref33,ref39}:
 \begin{equation}
   g^{+} \simeq 0.871, \hspace{.2in} g^{-} \simeq 0.174, \hspace{.2in} b \simeq 2.05,
 \end{equation}
and
 \begin{equation}
  g_{1}^{+} \simeq 0.610, \hspace{.2in} b_{1} \simeq 0.531.
 \end{equation}
The theoretical expression for the amplitude $\Gamma_{1}^{-}$ in Eq. (2.13) has not yet been evaluated. The values implied by the crossover Landau model for the corresponding universal amplitude ratios are
 \begin{equation}
  \Gamma_{0}^{+}/\Gamma_{0}^{-} = 5.0, \hspace{.2in} B_{1}/\Gamma_{1}^{+} = 0.87,
 \end{equation}
to be compared with the values $\Gamma_{0}^{+}/\Gamma_{0}^{-} = 4.95\pm 0.15$ and $B_{1}/\Gamma_{1}^{+} = 0.90\pm 0.21$ \cite{ref35}. A phenomenological reformulation of the two-term Landau model in terms of a parametric representation yielding slightly more accurate values for some universal amplitude ratios has been presented by Agayan {\em et al.} \cite{ref33}.

We define a crossover temperature $\tau_{\times}$ as
 \begin{equation}
  \tau_{\times} = (\Gamma_{1}^{+})^{-1/\Delta_{\mbox{\tiny s}}} = (g_{1}^{+})^{-1/\Delta_{\mbox{\tiny s}}} \frac{(\bar{u}\Lambda)^{2}}{c_{t}} (1-\bar{u})^{-1/\Delta_{\mbox{\tiny s}}},
 \end{equation}
or, equivalently,
 \begin{equation}
  t_{\times} = (g_{1}^{+})^{-1/\Delta_{\mbox{\tiny s}}}(\bar{u}\Lambda)^{2}(1-\bar{u})^{-1/\Delta_{\mbox{\tiny s}}}.
 \end{equation}
In the infinite-cutoff approximation ($\Lambda \rightarrow \infty$, $\bar{u}\rightarrow 0$, while $\bar{u}\Lambda$ remains finite), $\tau_{\times}$ becomes the so-called Ginzburg number \cite{ref22,ref31}. In general, the crossover behavior is governed by two system-dependent parameters, namely, the cutoff $\Lambda$ and the coupling constant $\bar{u}$. However, for the Ising model $\Lambda$ is a constant equal to unity independent of the range of interaction and only the coupling constant $\bar{u}$ governs the crossover behavior. One can also show that in the asymptotic crossover Landau model the appropriately scaled susceptibilities, $\tilde{\chi}^{\pm} \equiv \chi^{\pm} \tau_{\times}^{\gamma}/\Gamma_{0}^{\pm}$, and the scaled order parameter, $\tilde{\phi} \equiv \phi \tau_{\times}^{-\beta}/B_{0}$ become universal functions of a single argument $\tau/\tau_{\times} = t/t_{\times}$ \cite{ref22}.

After some algebra, one can obtain the following universal expression for the crossover behavior of the scaled susceptibility $\tilde{\chi}^{+}$ above $T_{\mbox{\scriptsize c}}$ in zero ordering field as a function of $\tau/\tau_{\times}$:
 \begin{eqnarray}
  \tilde{\chi}^{+} & = & \frac{(g_{1}^{+})^{\gamma -1}}{g^{+}} \left(\frac{\tau}{\tau_{\times}}\right)^{-1}\tilde{Y}^{(1-\gamma)/\Delta{\mbox{\tiny s}}} \nonumber \\
  & & \times \left[ 1 + \frac{u^{\ast}\nu}{2\Delta_{\mbox{\tiny s}}} \left( \frac{2 \tilde{Y}}{1-\tilde{Y}} + \frac{1}{\Delta_{\mbox{\tiny s}}} \right)^{-1} \right]^{-1},
 \end{eqnarray}
where the rescaled crossover function $\tilde{Y}$, can be obtained by solving
 \begin{equation}
  1 - \tilde{Y} = (g_{1}^{+}\tau/\tau_{\times})^{-1/2} \tilde{Y}^{(1-\nu)/2\Delta_{\mbox{\tiny s}}}.
 \end{equation}
The scaled susceptibility $\tilde{\chi}^{-}$ and the scaled order parameter $\tilde{\phi}$ below $T_{\mbox{\scriptsize c}}$ are to be obtained by imposing the condition $(\partial A/ \partial M)_{t} = 0$ for the phase boundary and solving Eqs. (2.5)-(2.9) numerically.

We note that the crossover Landau model as specified above is an asymptotic version of the crossover Landau model \cite{ref22,ref31} valid for $\Lambda/\kappa \gg 1$. This asymptotic version appears to be adequate for the temperature ranges covered by the numerical studies of Luijten and Binder.

The crossover Landau model has been derived from the renormalization-group theory of critical phenomena by using the technique of renormalization-group matching \cite{ref13,ref40,ref41}. In this approach one considers the relation between $\Delta A$ at a given cutoff $\Lambda$ and $\Delta A$ at another cutoff $\Lambda e^{-l}$, where $l$ is a variable. One then tries to determine a special ``matching-point'' value $l=l^{\ast}$ for which $\Delta A$ reduces to the known classical expression in the absence of fluctuations. One limitation is that the value of $l^{\ast}$ that satisfies the matching condition for $\Delta A$ will not exactly satisfy the matching condition for its derivatives like $\chi$. Secondly, for practical reasons $l^{\ast}$ has been selected to satisfy the matching condition $M=0$ \cite{ref20,ref21}. A comparison with the numerical data for the three-dimensional Ising model will enable us to make an assessment of the quality of these approximations.

\section{Application to the Three-dimensional Ising Model}
In the Monte Carlo simulations a three-dimensional equivalent-neighbor model was adopted for which the Hamiltonian ${\cal H}$ is given by \cite{ref11}
 \begin{equation}
  {\cal H}/k_{B}T = -\sum_{\langle ij \rangle}K({\bf r}_{i}-{\bf r}_{j})s_{i}s_{j}, \label{hamil}
 \end{equation}
where $k_{\mbox{\scriptsize B}}$ is Boltzmann's constant, $s_{i}=\pm 1$, the sum runs over all spin pairs, and the spin-spin coupling is defined as $K({\bf r})=J>0$ for $|{\bf r}| \leq R_{\mbox{\scriptsize m}}$ and $K({\bf r}) = 0$ for $|{\bf r}| > R_{\mbox{\scriptsize m}}$. However, in a renormalization-group analysis, various critical properties have been obtained as functions of an effective interaction range $R$ defined by
 \begin{eqnarray}
  R^{2} & \equiv & \sum_{j\neq i}({\bf r}_{i}-{\bf r}_{j})^{2}K_{ij}/\sum_{j\neq i}K_{ij} \nonumber \\
  & = &\frac{1}{q}\sum_{j\neq i}|{\bf r}_{i}-{\bf r}_{j}|^{2} \hspace{.2in} \mbox{with} \hspace{.1in} |{\bf r}_{i}-{\bf r}_{j}|\leq R_{m}, 
 \end{eqnarray}
where $K_{ij}$ stands for $K({\bf r}_{i}-{\bf r}_{j})$ while $q$ is the coordination number. In the limit of an infinite interaction range $R\rightarrow \infty$, $R$ is related to $R_{\mbox{\scriptsize m}}$ by $R^{2} = 3R_{\mbox{\scriptsize m}}^{2}/5$, while $R = R_{\mbox{\scriptsize m}}$ for $R_{\mbox{\scriptsize m}} =1$. The dependence of the effective interaction range $R$ on $R_{\mbox{\scriptsize m}}$ is shown in Fig. 1. Also shown is an approximate relation \cite{ref42}
 \begin{equation}
  R^{2} = \mbox{$\frac{3}{5}$}R_{\mbox{\scriptsize m}}^{2}(1+\mbox{$\frac{2}{3}$}R_{\mbox{\scriptsize m}}^{-2})
 \end{equation}
encompassing the known behavior in the limits $R\rightarrow \infty$ and $R=1$.

In order to apply the theoretical crossover model to the three-dimensional Ising model we need to specify the system-dependent constants. From the mean-field theory of the Ising model one obtains for the coefficients in the classical expansion (2.1) \cite{ref43}
 \begin{equation}
  a_{0} = 1, \hspace{.3in} u_{0} = 2.
 \end{equation}
The physical cutoff wave number for the Ising model is $\pi/a$, so that
 \begin{equation}
  \Lambda = 1.
 \end{equation}
In a previous paper \cite{ref42}, $\Lambda$ of the three-dimensional Ising model was taken to be $\pi$. However, we have subsequently realized that the spatial derivatives in the gradient term are already scaled by $\pi/a$.

A renormalization-group analysis of Eq. (3.1) has shown that $\bar{u} \propto R^{-4}$ for large $R$ \cite{ref7,ref11}. It then follows from Eq. (2.3) that $c_{t} \propto R^{-2}$ and $c_{\rho} \propto R$ and we write
 \begin{equation}
  \bar{u} = \frac{\bar{u}_{0}}{R^{4}}, \hspace{.2in} c_{t} = \frac{c_{t0}}{R^{2}}, \hspace{.2in} c_{\rho} = c_{t}^{-1/2} = R c_{t0}^{-1/2}.
 \end{equation}
In a previous attempt to compare the crossover model with the numerical results for the three-dimensional Ising model, $\bar{u}_{0}$ and $c_{t0}$ were treated as adjustable constants \cite{ref42}. In the present paper we shall adopt a priori estimates for $\bar{u}_{0}$ and $c_{t0}$. It has been suggested by Luijten \cite{ref11} that $\bar{u}_{0}$, and consequently $c_{t0}$, may actually exhibit a remaining weak dependence on $R$ for small $R$. From Eqs. (2.3), (3.4) and (3.5) we note that the coefficients $\bar{u}_{0}(R)$ and $c_{t0}(R)$ are related by
 \begin{equation}
  c_{t0}^{2}(R) = \mbox{$\frac{1}{2}$}u^{\ast} \bar{u}_{0}(R).
 \end{equation}
The values of the coefficients $\bar{u}_{0}$ and $c_{t0}$ can readily be determined for large $R$. For this purpose we note that the amplitude $\bar{\xi}_{0}^{+}$ of the power law $\bar{\xi} = \bar{\xi}_{0}^{+} \tau^{-1/2}$ for the  classical correlation length $\bar{\xi}$ is given by $\bar{\xi}_{0}^{+} = (c_{0}/a_{0})^{1/2}$ \cite{ref22}. Starting from an expression given by Fisher and Burford \cite{ref44} and extending it to arbitrary interaction ranges, one finds that $\bar{\xi}_{0}^{+}$ becomes equal to $R/\sqrt{2d}$, so that $c_{0}$ becomes equal to $R^{2}/2d$ for large $R$. Using Eqs. (2.3), (3.4) and (3.6) we thus obtain for large $R$:
 \begin{equation}
  c_{t0}(\infty) = 6/\pi^{2} \simeq 0.608, \hspace{.2in} \bar{u}_{0}(\infty) = \frac{2}{u^{\ast}}c_{t0}^{2}(\infty) \simeq 1.566.
 \end{equation}

In the present paper we shall use two approximations for the coupling constant $\bar{u}$ of the three-dimensional Ising model. One estimate is obtained by assuming that $\bar{u}$ remains proportional to $R^{-4}$ for all values of $R$:
 \begin{equation}
  \bar{u} = \frac{\bar{u}_{0}(\infty)}{R^{4}}.
 \end{equation}
Using Eq. (3.9) in conjunction with Eqs. (3.6)-(3.8) we can then predict from the crossover Landau model the amplitude $\Gamma_{1}^{+}$ which is related to the crossover temperature $\tau_{\times}$ through Eq. (2.23). The resulting values for $\Gamma_{1}^{+}$ as a function of $R^{3}$ are represented by the solid curve in Fig. 2. We note that, since $\Delta_{\mbox{\scriptsize s}} \simeq 1/2$, $\Gamma_{1}^{+}$ will approximately vary as $R^{3}$ for large $R$.

A second estimate for $\Gamma_{1}^{+}$ is obtained from numerical values calculated by Luijten for the amplitude $b_{1}(R)$ of the correction-to-scaling contribution to the finite size effects \cite{ref11}. For large values of $R$ where $\bar{u}$ approaches zero, $b_{1}(R)$ becomes proportional to $\Gamma_{1}^{+}(R)R^{\alpha\gamma/\Delta_{\mbox{\tiny s}}}$ so that we may write
 \begin{equation}
  \Gamma_{1}^{+}(R) = b_{0}(R)b_{1}(R)R^{-\alpha\gamma/\Delta_{\mbox{\tiny s}}},
 \end{equation}
where $b_{0}(R)$ is a coefficient with a finite limiting value
 \begin{equation}
  b_{0}(\infty) = -2.6 \pm 0.5.
 \end{equation}
Values deduced for $\Gamma_{1}^{+}$ from the numerical data obtained by Luijten for $b_{1}(R)$, when the coefficient $b_{0}(\infty)$ in Eq. (3.10) is approximated by $b_{0}(\infty)$ for all $R$, are indicated by the circles in Fig. 2. These values can be reproduced from Eq. (2.23) for the crossover Landau model if we adopt the following approximation for $\bar{u}$:
 \begin{equation}
  \bar{u} = \frac{\bar{u}_{0}(\infty)}{R^{4}} ( 1 + 0.9R^{-2}),
 \end{equation}
as shown in the dotted curve in Fig. 2.

Equations (3.9) and (3.12) are the two approximations that will be considered in applying the crossover Landau model to the Ising lattice. As we shall see, the predicted critical crossover behavior of the susceptibility is not very sensitive to these different choices for the coupling constant $\bar{u}$ for most values of $R$.

\section{Results of Analysis}
In Fig. 3 we present a plot of the scaled susceptibility $\tilde{\chi} = \chi^{+} \tau_{\times}^{\gamma}/ \Gamma_{0}^{+}$ above $T_{\mbox{\scriptsize c}}$ as a function of $\tau/\tau_{\times}$. The solid curve represents the values predicted from Eq. (2.26) for the crossover Landau model with $\bar{u} = \bar{u}_{0}(\infty)R^{-4}$ in accordance with Eq. (3.9). The symbols indicate the numerical values obtained by Luijten and Binder \cite{ref10,ref11}. A sensitive test of the crossover behavior is obtained by considering an effective susceptibility exponent $\gamma_{\mbox{\scriptsize eff}}^{+}$ defined by \cite{ref45}:
 \begin{equation}
  \gamma_{\mbox{\scriptsize eff}}^{+} = -d\log \tilde{\chi}/d \log\tau \hspace{.2in} (\tau > 0).
 \end{equation}
Figure 4 shows the effective exponent $\gamma_{\mbox{\scriptsize eff}}^{+}$ as a function of $\tau/ \tau_{\times}$. The symbols indicate values deduced by numerical differentiation of the susceptibilities obtained by Luijten \cite{ref11}. The solid curve represents the effective exponent values predicted from the crossover model. Note that this exponent is related to the third derivative of the Helmholtz-energy density $\Delta \tilde{A}$. In Fig. 5 we show again $\gamma_{\mbox{\scriptsize eff}}^{+}$ as a function of $\tau/ \tau_{\times}$ but now with $\bar{u} = \bar{u}_{0}(\infty)R^{-4} (1+0.9R^{-2})$ in accordance with Eq. (3.12). We remark that the theoretical prediction has a possible inaccuracy at the percentage level, since the values adopted for the various universal constants have a finite accuracy. In addition, we have only limited theoretical information for the values of $\bar{u}$ for small $R$. Based on the currently available theoretical information we conclude from Figs. 4 and 5 that the crossover Landau model yields a satisfactory representation of the crossover behavior of the susceptibility above $T_{\mbox{\scriptsize c}}$ for the three-dimensional Ising model with varying interaction ranges.

As mentioned in Section 2, the crossover Landau model has been derived from the renormalization-group theory of critical phenomena by renormalization-group matching, but the match-point condition actually adopted corresponds to $M=0$ \cite{ref20}. This is correct for the susceptibility above $T_{\mbox{\scriptsize c}}$ at $M=0$, but represents an approximation below $T_{\mbox{\scriptsize c}}$ where $M_{\mbox{\scriptsize cxc}} \neq 0$. Hence, we expect the crossover Landau model to be less accurate for a description of the critical crossover behavior of the susceptibility and of the order parameter below $T_{\mbox{\scriptsize c}}$.

In Fig. 6 we show the scaled order parameter $\tilde{\phi} = \phi \tau_{\times}^{-\beta}/B_{0}$ and in Fig. 7 the scaled susceptibility $\tilde{\chi}^{-} = \chi^{-} \tau_{\times}^{\gamma}/\Gamma_{0}^{-}$ as a function of $-\tau/\tau_{\times}$. The solid curves represent again the values predicted from the crossover Landau model and the symbols indicate again the numerical data obtained by Luijten and Binder \cite{ref10}. In both cases the asymptotic Eq. (3.9) was adopted for $\bar{u}$. While the actual differences with the numerical values for $\phi$ and $\chi^{-}$ are small and probably within the accuracy these properties can be measured experimentally for real systems, the deviations are systematic. This becomes evident when one takes the logarithmic derivative so as to obtain the effective order-parameter exponent
 \begin{equation}
  \beta_{\mbox{\scriptsize eff}} = d\log \phi/ d\log |\tau|, \hspace{.2in} (\tau < 0)
 \end{equation}
shown in Fig. 8 and the effective susceptibility exponent $\gamma_{\mbox{\scriptsize eff}}^{-}$
 \begin{equation}
  \gamma_{\mbox{\scriptsize eff}}^{-} = -d\log \chi /d \log |\tau|, \hspace{.2in} (\tau < 0)
 \end{equation}
shown in Fig. 9. The effective exponents predicted by the crossover Landau model as a function of $\tau/\tau_{\times}$ have the correct shape but the location of the crossover temperature $\tau = \tau_{\times}$ is shifted from that implied by the numerical data. This shift amounts to about $1/2$ of a decade for $\beta_{\mbox{\scriptsize eff}}$ and $2/3$ of a decade for $\gamma_{\mbox{\scriptsize eff}}^{-}$. The shift is larger for $\gamma_{\mbox{\scriptsize eff}}^{-}$ than for $\beta_{\mbox{\scriptsize eff}}$, since $\gamma_{\mbox{\scriptsize eff}}^{-}$ is related to the third derivative and $\beta_{\mbox{\scriptsize eff}}$ is related to the second derivative of the crossover function deduced for $\Delta A$ by renormalization-group matching.

Crossover Landau models with improved approximants for the rescaling functions ${\cal T}$, ${\cal D}$, ${\cal U}$, and ${\cal K}$ have been considered by Tang {\em et al.} \cite{ref32}, but they never have been used for practical use because of increased complexity.

The solid curves in Figs. 4, 5, 8 and 9 actually represent the effective exponents for $R_{\mbox{\scriptsize m}} > 1$ where $\bar{u} < 1$. For $R_{\mbox{\scriptsize m}} = R = 1$ the leading amplitudes $\Gamma_{0}^{\pm}$ and $B_{0}$ have been determined by Liu and Fisher \cite{ref35} from a careful analysis of series expansions. The value of $\Gamma_{0}^{+} = 1.095$ corresponding to $\gamma = 1.239$ implies $\bar{u} = 1.17$ as follows from Eq. (2.15). For $\bar{u} > 1$, the correction-to-scaling amplitudes $\Gamma_{1}^{+}$ and $B_{1}$ become negative in agreement with an earlier observation of Liu and Fisher \cite{ref46}. A comparison of the amplitudes calculated from the crossover model for $\bar{u} = 1.17$ with the amplitudes deduced by Liu and Fisher \cite{ref35} from series expansions is presented in Table I. For $\bar{u} > 1$, the asymptotic value of the effective susceptibility exponent $\gamma_{\mbox{\scriptsize eff}}^{\pm}$ is approached from above and that of the effective order-parameter exponent $\beta_{\mbox{\scriptsize eff}}$ from below \cite{ref33,ref47}. The dependence of the effective exponents on $\tau/\tau_{\times}$ with $\tau_{\times} \equiv |\Gamma_{1}^{+}|^{-1/\Delta_{\mbox{\tiny s}}}$ for the case $R=1$, corresponding to $\bar{u} = 1.17$, is represented by the dashed curves in Figs. 4, 5, 8 and 9 in the temperature range covered by the numerical simulations.

From Fig. 9 we note that the numerical data as well as the crossover Landau model show that $\gamma_{\mbox{\scriptsize eff}}^{-}$ passes through a minimum as a function of $\tau/\tau_{\times}$ before reaching its mean-field value unity, (see Ref. 48). An interesting question, raised earlier by Fisher \cite{ref49}, is whether also $\gamma_{\mbox{\scriptsize eff}}^{+}$ can assume values smaller than unity. Within the accuracy of our analysis, we do not find any evidence that the effective susceptibility exponent above $T_{\mbox{\scriptsize c}}$ can assume values smaller than unity, as was already discussed in Ref. 10.

\section{Discussion}
In the present paper we have applied a crossover Landau model, previously derived for the representation of thermodynamic properties of fluids near the critical point to predict the critical crossover behavior of the susceptibility above $T_{\mbox{\scriptsize c}}$ and of the order parameter and susceptibility below $T_{\mbox{\scriptsize c}}$ for the three-dimensional Ising model with various interaction ranges. A comparison with numerical data, recently obtained by Luijten and Binder \cite{ref10,ref11}, demonstrates that the crossover Landau model yields a good representation of the crossover behavior of the susceptibility above $T_{\mbox{\scriptsize c}}$. Below $T_{\mbox{\scriptsize c}}$ there are some systematic differences from the numerical data which are small for $\phi$ and $\chi^{-}$ themselves, but which become more pronounced when one considers the corresponding effective exponents. In applications of the crossover model to real fluids, residual deviations are partially accounted for by using $\bar{u}$ and $\Lambda$, as well as the coefficients $a_{0}$ and $u_{0}$ in the classical expansion (2.1) of the Helmholtz-energy density, or, equivalently, $c_{t}$ and $c_{\rho}$ (cf. Eq. (2.3)) as adjustable coefficients \cite{ref42}.

The analysis presented in this paper is based on an asymptotic version of the crossover Landau model which deals with the crossover from fluctuation-induced critical behavior to asymptotic mean-field critical behavior. For some other applications one needs to consider a more general version of the crossover Landau model that is obtained when the ratio $\Lambda/\kappa$ in Eq. (2.10) is replaced by $(1+\Lambda^{2}/\kappa^{2})^{1/2}$ \cite{ref22}. This generalization was originally introduced to account for crossover to nonasymptotic mean-field behavior in simple fluids \cite{ref21}. Subsequently, it has become evident that this generalization becomes essential in the case of complex fluids, when the crossover to mean-field behavior does not result from long-range interactions, but from a decrease of the effective cutoff wave number $\Lambda$. In that case $\Lambda^{-1}$ plays the role of a correlation length associated with an additional order parameter. Then the mean-field limit corresponds to mean-field tricriticality, when both $\xi$ and $\Lambda^{-1}$ diverge \cite{ref50,ref51}.

\acknowledgements
The authors have benefited from many stimulating discussions with M. E. Fisher, to whom this article is dedicated with great appreciation. Kh. S. Abdulkadirova and V. A. Agayan provided assistance in the initial stage of this research.

The research of M. A. Anisimov and J. V. Sengers is supported by the Chemical Sciences, Geosciences and Biosciences Division, Office of Basic Energy Sciences, Office of Science, Department of Energy under Grant No. DE-FG02-95-ER-14509. The research of Y. C. Kim and E. Luijten have been supported by Grant No. CHE99-81772 from the National Science Foundation to M. E. Fisher at the University of Maryland.

\pagebreak

\begin{table}[ht]
\caption{The critical amplitudes of the susceptibilities above and below $T_{\mbox{\scriptsize c}}$ and of the order parameter from the crossover two-term Landau model (CLM) using $\bar{u}=1.17$, and from series expansions [35] for $R=1$.}
\vspace{.2in}
\begin{tabular}{|c|c|c|c|c|c|} 
      & $\Gamma_{0}^{+}$ & $\Gamma_{0}^{-}$ & $B_{0}$ & $\Gamma_{1}^{+}$ & $B_{1}$  \\ \hline\hline
$\;$ CLM $\;$ &  1.095 $\;$ &  0.219 $\;$ & 1.71 $\;$ & $-0.0636$ $\;$ &  $-0.055$ $\;$  \\ \hline
$\;$ Series exp.  $\;$ & 1.095 $\;$ & 0.220 $\;$ & 1.66 $\;$  &  --- $\;$  & ---  $\;$  \\ \hline\hline
      &  $\Gamma_{0}^{+}/\Gamma_{0}^{-}$ $\;$ & $B_{1}/\Gamma_{1}^{+}$ $\;$ \\ \hline\hline
$\;$ CLM $\;$ &  5.0 $\;$ & 0.87 $\;$ \\ \hline
$\;$ Series exp.  $\;$ &  4.98 $\;$ & --- $\;$ \\
\end{tabular}

\end{table}
\begin{figure}[ht]

\centerline{\epsfig{figure=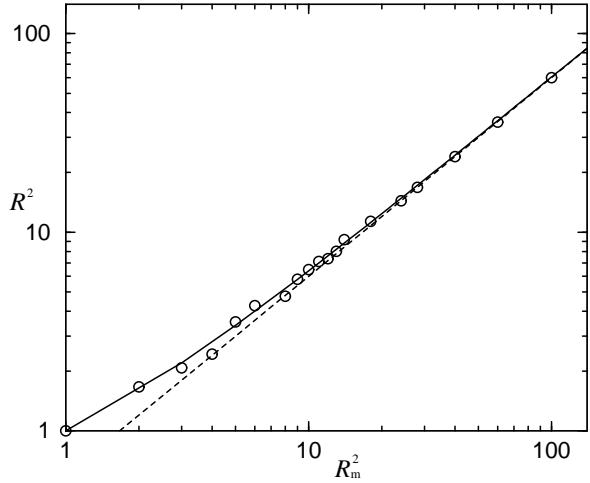,width=3in,angle=-90}}
\vspace{.2in}
 \caption{Square of the effective interaction range $R$ as a function of $R_{\mbox{\scriptsize m}}^{2}$. The symbols indicate values calculated by Luijten [11]. The solid curve represents the approximant given by Eq. (3.3). The dotted line represents the asymptotic relation $R^{2} = 3 R_{\mbox{\scriptsize m}}^{2}/5$ for large $R$.}

\end{figure}
\begin{figure}[ht]
\centerline{\epsfig{figure=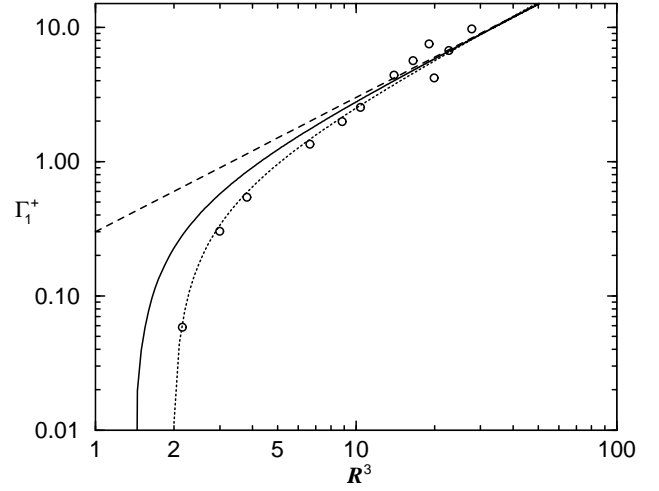,width=3in,angle=-90}}
\vspace{.2in}
 \caption{Amplitude $\Gamma_{1}^{+}$ of the correction-to-scaling contribution in the expansion (2.12) for the susceptibility. The solid curve represent values predicted from the crossover Landau model with $\bar{u} = \bar{u}_{0}(\infty) R^{-4}$ and the dotted curve with $\bar{u} = \bar{u}_{0}(\infty) R^{-4}(1 + 0.9R^{-2})$. The circles indicate numerical estimates deduced from finite-size effects obtained by Luijten [11]. The dashed lines represents the asymptotic linear dependence of $\Gamma_{1}^{+}$ on $R^{3}$.}

\end{figure}
\begin{figure}[ht]
\centerline{\epsfig{figure=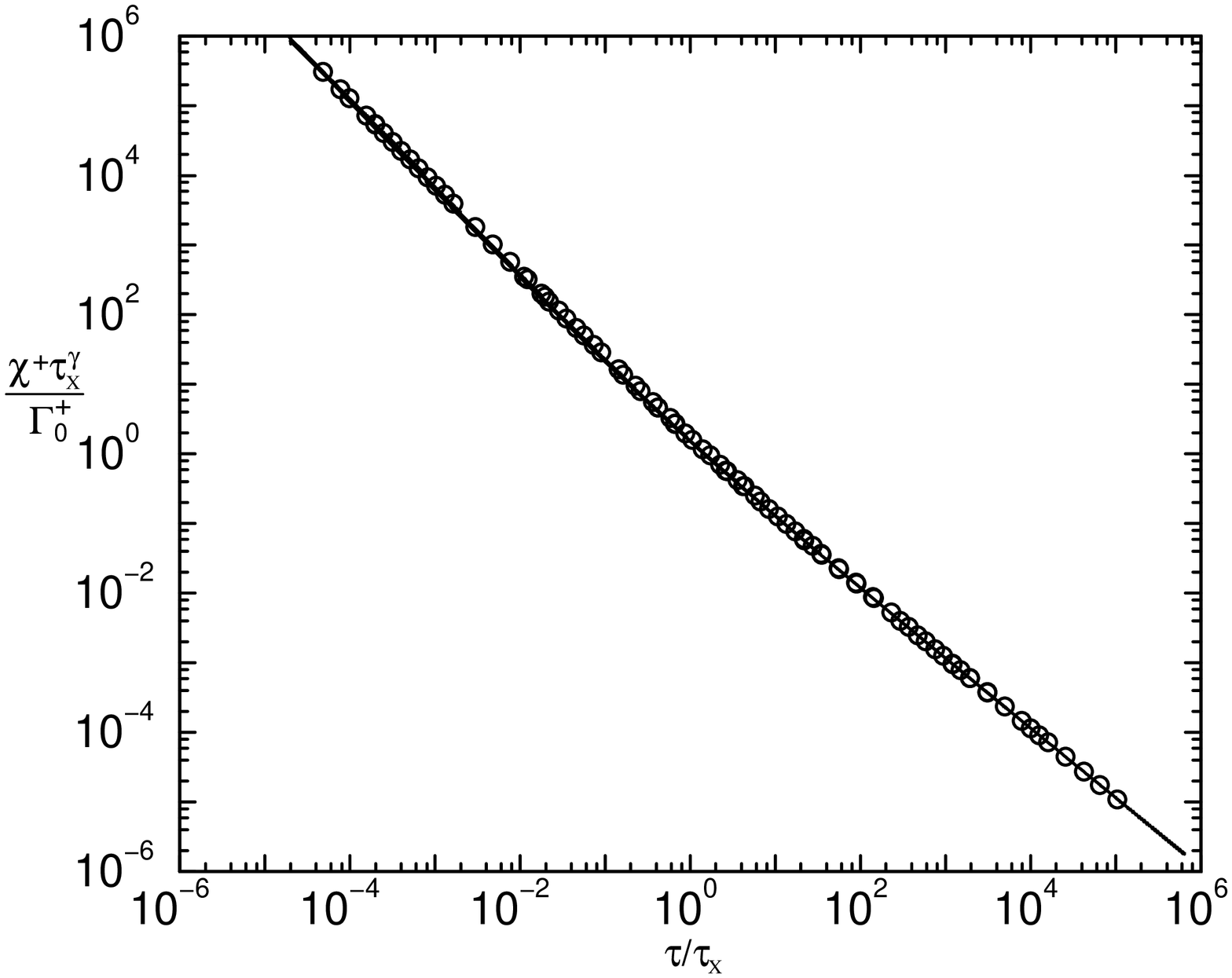,width=3in,angle=-90}}
\vspace{.2in}
 \caption{Scaled susceptibility $\tilde{\chi}^{+} = \chi^{+}\tau_{\times}^{\gamma} \Gamma_{0}^{+}$ as a function of $\tau/\tau_{\times}$. The symbols indicate numerical values from the Monte Carlo simulations [10,11]. The curve represents values calculated from the crossover Landau model with $\bar{u} = \bar{u}_{0}(\infty) R^{-4}$.}

\end{figure}
\begin{figure}[ht]
\centerline{\epsfig{figure=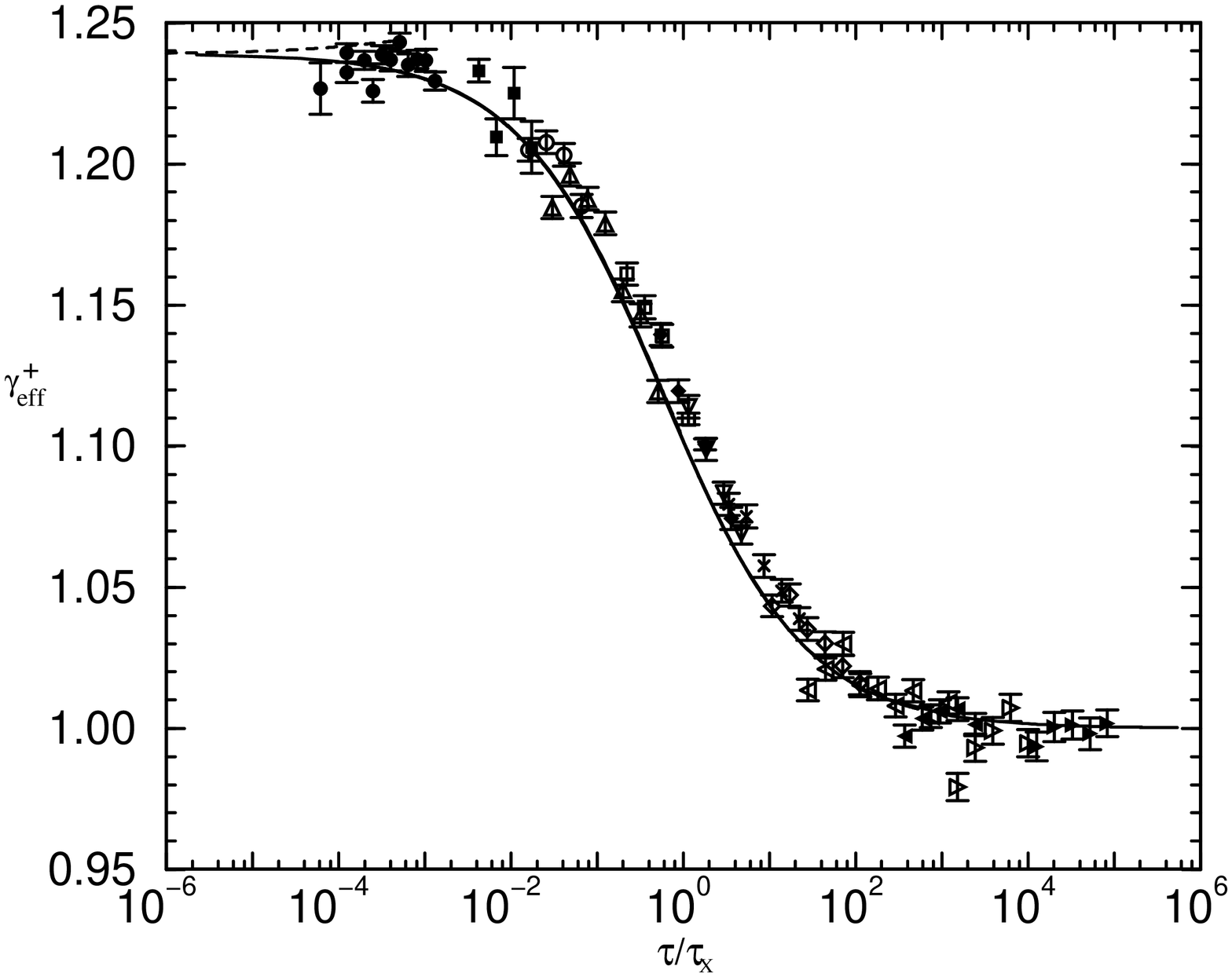,width=3in,angle=-90}}
\vspace{.2in}
 \caption{Effective susceptibility exponent $\gamma_{\mbox{\scriptsize eff}}^{+}$ as a function of $\tau/\tau_{\times}$. The symbols indicate numerical values deduced by numerical differentiation of the results from the Monte Carlo simulations [11]. The solid curve represents values calculated from the crossover Landau model with $\bar{u}=\bar{u}_{0}(\infty)R^{-4}$ for $R_{\mbox{\scriptsize m}} \geq 2$. The dashed curve represents $\gamma_{\mbox{\scriptsize eff}}^{+}$ for $R_{\mbox{\scriptsize m}} = 1$ corresponding to $\bar{u} > 1$.}

\end{figure}
\begin{figure}[ht]
\centerline{\epsfig{figure=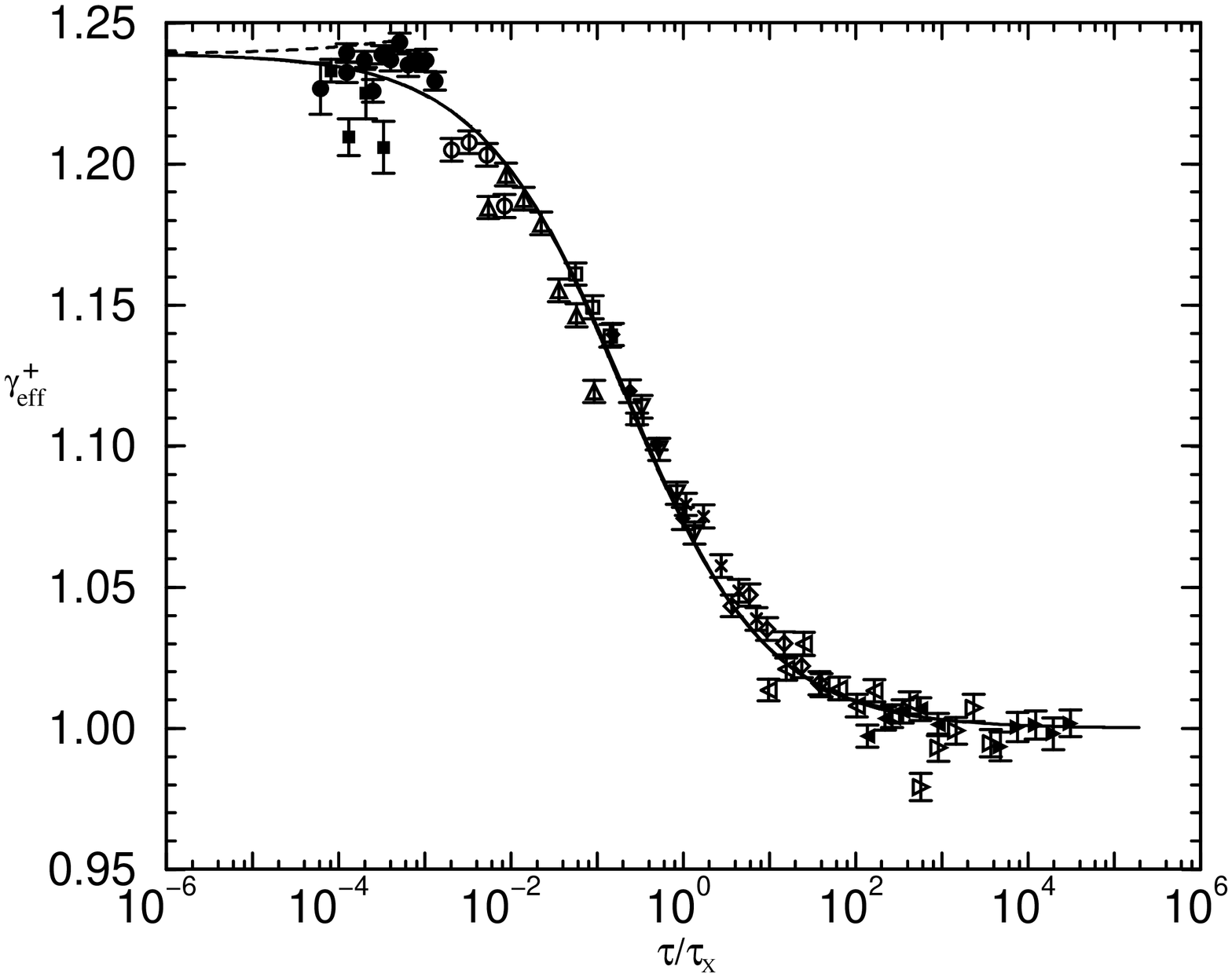,width=3in,angle=-90}}
\vspace{.2in}
 \caption{Effective susceptibility exponent $\gamma_{\mbox{\scriptsize eff}}^{+}$ as a function of $\tau/\tau_{\times}$. The symbols indicate values deduced by numerical differentiation of the results from the Monte Carlo simulations as in Fig. 4. The solid curve represents values calculated from the crossover Landau model with $\bar{u} = \bar{u}_{0}(\infty)R^{-4} (1+0.9R^{-2})$ for $R_{\mbox{\scriptsize m}} \geq 2$. The dashed curve represents $\gamma_{\mbox{\scriptsize eff}}^{+}$ for $R_{\mbox{\scriptsize m}} = 1$ corresponding to $\bar{u} > 1$.}

\end{figure}
\begin{figure}[ht]
\centerline{\epsfig{figure=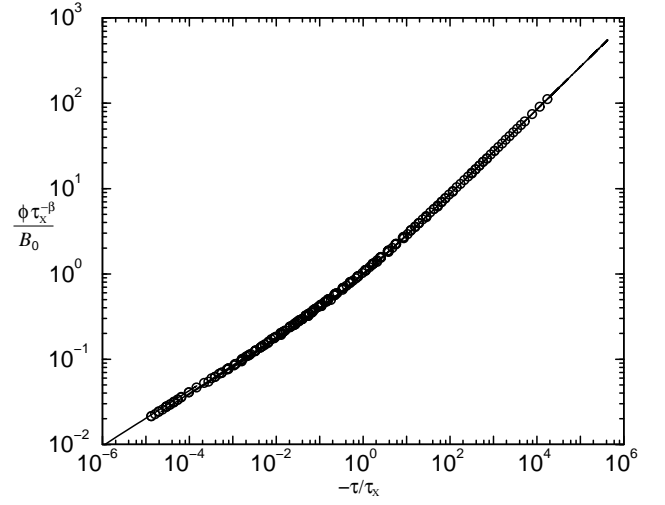,width=3in,angle=-90}}
\vspace{.2in}
 \caption{Scaled order-parameter $\tilde{\phi} = \phi \tau_{\times}^{-\beta}/B_{0}$ as a function of $-\tau/\tau_{\times}$. The symbols indicate numerical values from the Monte Carlo simulations [10,11]. The curve represents values calculated from the crossover Landau model with $\bar{u} = \bar{u}_{0}(\infty) R^{-4}$.}

\end{figure}
\begin{figure}[ht]
\centerline{\epsfig{figure=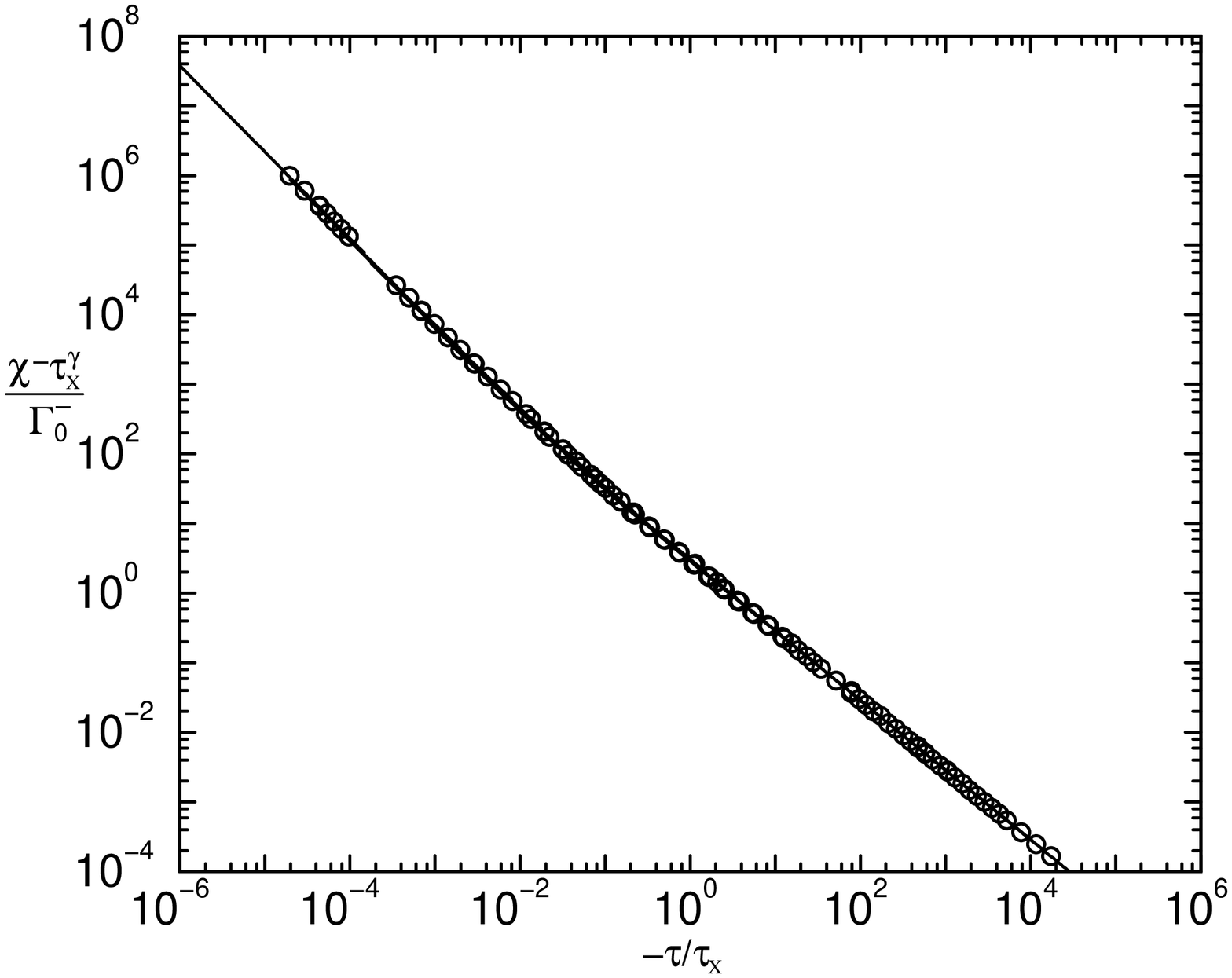,width=3in,angle=-90}}
\vspace{.2in}
 \caption{Scaled susceptibility $\tilde{\chi}^{-} = \chi^{-}\tau_{\times}^{\gamma}/\Gamma_{0}^{-}$ as a function of $-\tau/\tau_{\times}$. The symbols indicate numerical values from the Monte Carlo simulations [10,11]. The curve represents values calculated from the crossover model with $\bar{u}= \bar{u}_{0}(\infty)R^{-4}$.}

\end{figure}
\begin{figure}[ht]
\centerline{\epsfig{figure=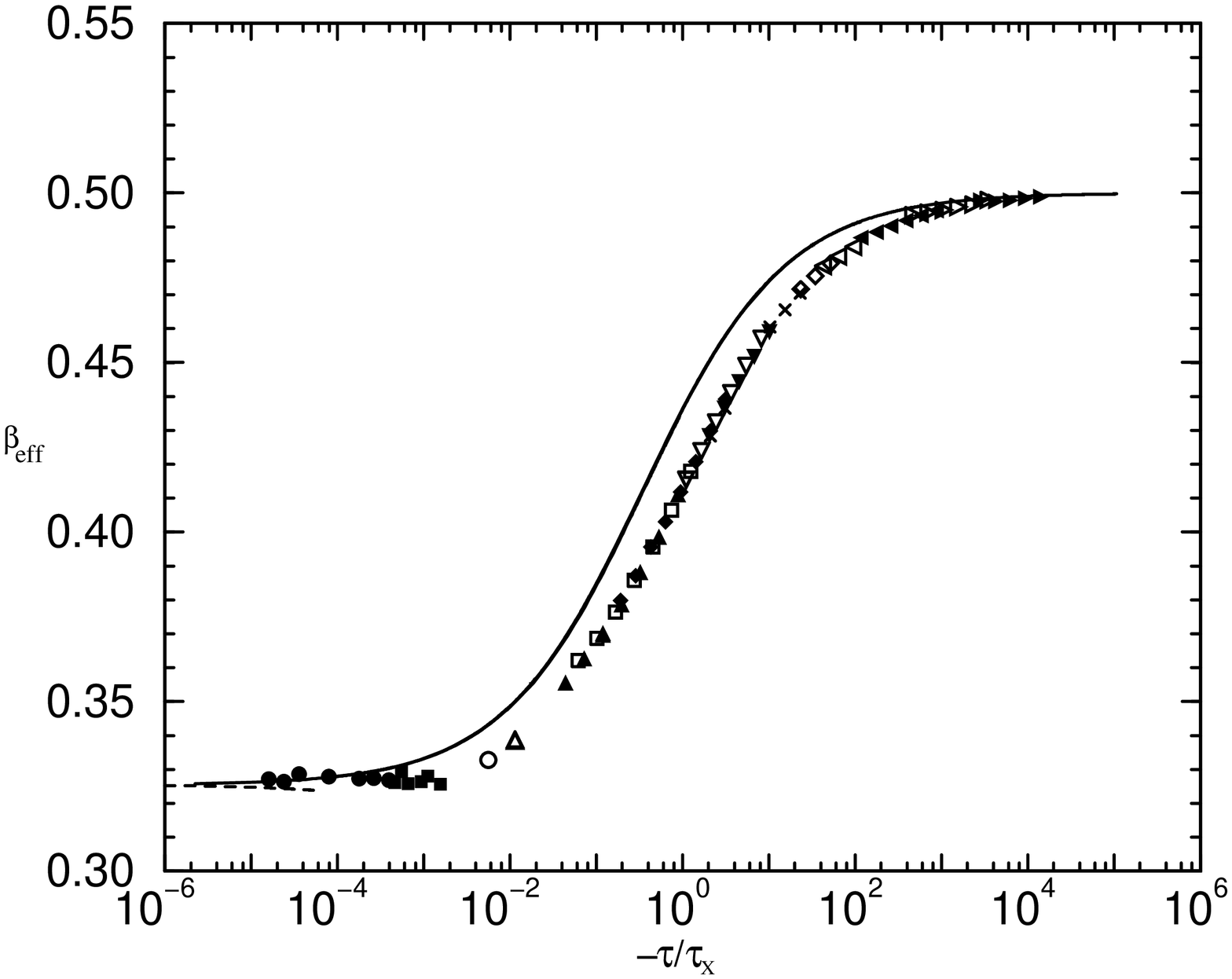,width=3in,angle=-90}}
\vspace{.2in}
 \caption{Effective order-parameter exponent $\beta_{\mbox{\scriptsize eff}}$ as a function of $-\tau/\tau_{\times}$. The symbols indicate values deduced by numerical differentiation of the results from the Monte Carlo simulations [11]. The solid curve represents values calculated from the crossover Landau model with $\bar{u} = \bar{u}_{0}(\infty) R^{-4}$ for $R_{\mbox{\scriptsize m}} \geq 2$. The dashed curve represents $\beta_{\mbox{\scriptsize eff}}$ for $R_{\mbox{\scriptsize m}} = 1$ corresponding to $\bar{u} >1$.}

\end{figure}
\begin{figure}[ht]
\centerline{\epsfig{figure=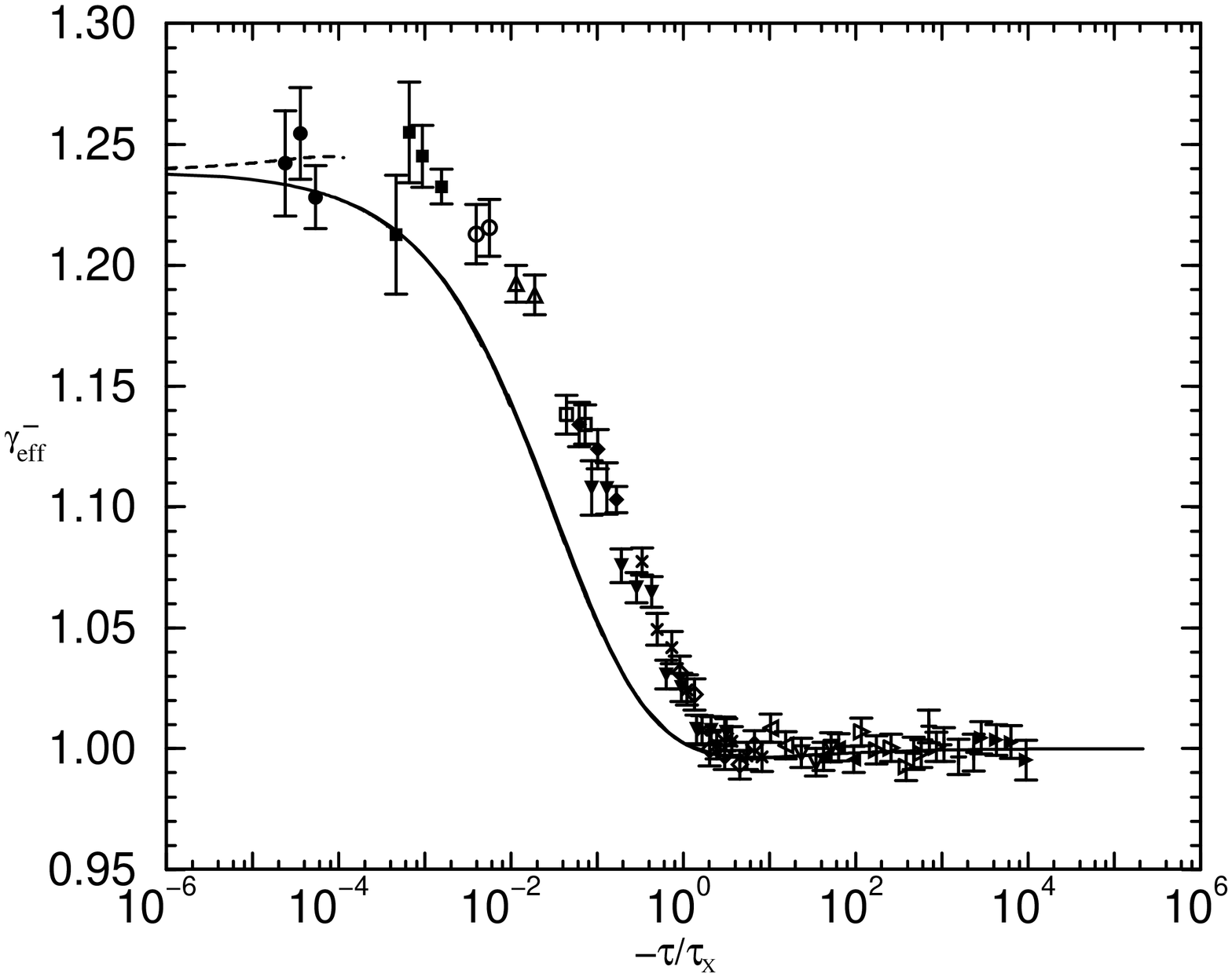,width=3in,angle=-90}}
\vspace{.2in}
 \caption{Effective susceptibility exponent $\gamma_{\mbox{\scriptsize eff}}^{-}$ as a function of $-\tau/\tau_{\times}$. The symbols indicate values deduced by numerical differentiation of the results from the Monte Carlo simulations [11]. The solid curve represents values calculated from the crossover Landau model with $\bar{u} = \bar{u}_{0}(\infty)R^{-4}$ for $R_{\mbox{\scriptsize m}} \geq 2$. The dashed curve represents $\gamma_{\mbox{\scriptsize eff}}^{-}$ for $R_{\mbox{\scriptsize m}} = 1$ corresponding to $\bar{u} > 1$.}

\end{figure}

\end{document}